\documentclass[12pt]{iopart}
\usepackage{graphicx}

\usepackage{iopams}
\newcommand{\OLo}{\Omega_{\Lambda}^0}

\newcommand{\rmr}{\rho_m}
\newcommand{\pmr}{p_m}

\newcommand{\wm}{\omega_m}

\newcommand{\amr}{\alpha_m}

\newcommand{\rL}{\rho_{\CC}}

\newcommand{\rLo}{\rho_{\CC}^0}

\newcommand{\CC}{\Lambda}
\newcommand{\CCo}{\Lambda_0}

\newcommand{\f}{\tilde{f}}
\newcommand{\cM}{{\cal M}}
\newcommand{\cMd}{{\cal M}^2}

\begin{document}
\title{Dark energy:
a quantum fossil from the inflationary Universe?}

\author{Joan Sol\`{a}\,
}

\address{High Energy Physics Group, Dept. ECM, and Institut de Ciencies del Cosmos, Universitat de Barcelona,
Diagonal 647, 08028 Barcelona, Catalonia, Spain}

\eads{sola@ifae.es}

\begin{abstract}
The discovery of dark energy (DE) as the physical cause for the
accelerated expansion of the Universe is the most remarkable
experimental finding of modern cosmology. However, it leads to
insurmountable theoretical difficulties from the point of view of
fundamental physics. Inflation, on the other hand, constitutes
another crucial ingredient, which seems necessary to solve other
cosmological conundrums and provides the primeval quantum seeds for
structure formation. One may wonder if there is any deep
relationship between these two paradigms. In this work, we suggest
that the existence of the DE in the present Universe could be linked
to the quantum field theoretical mechanism that may have triggered
primordial inflation in the early Universe. This mechanism, based on
quantum conformal symmetry, induces a logarithmic,
asymptotically-free, running of the gravitational coupling. If this
evolution persists in the present Universe, and if matter is
conserved, the general covariance of Einstein's equations demands
the existence of dynamical DE in the form of a running cosmological
term, $\CC$, whose variation follows a power law of the redshift.

\end{abstract}

\section{Introduction}

Modern Cosmology incorporates the notion of dark energy (DE) as an
experimental fact that accounts for the physical explanation of the
observed accelerated expansion\,\cite{SNe,WMAP3Y}. Although the
nature of the DE is not known, one persistent possibility is the
90-years-old cosmological constant (CC) term, $\CC$, in Einstein's
equations. In recent times, one is tempted to supersede this
hypothesis with another, radically different, one: viz. a slowly
evolving scalar field $\phi$ (``quintessence'') whose potential,
$V(\phi)\gtrsim 0$, could explain the present value of the DE and
whose equation of state (EOS) parameter $\omega_{\phi}=
p_{\phi}/\rho_{\phi}\simeq -1+\dot\phi^2/V(\phi)$ is only slightly
larger than $-1$ (hence insuring a negative pressure mimicking the
$\CC$ case)\,\cite{weinberg}. The advantage to think this way is
that the DE can then be a dynamical quantity taking different values
throughout the history of the Universe. However, this possibility
can not explain why the DE is entirely due to such an \textit{ad
hoc} scalar field and why the contributions to the vacuum energy
from the other fields (e.g. the electroweak Standard Model ones)
must not be considered. In short, it does not seem to be such a
wonderful idea to invent the field $\phi$ and simply replace
$\rL=\CC/8\pi\,G$ (the energy density associated to $\CC$, where $G$
is Newton's constant) with $\rho_{\phi}\simeq V(\phi)$. One has to
explain, too, why the various contributions (including the
additional one $V(\phi)$!) must conspire to generate the tiny value
of the DE density at present -- the ``old CC
problem''\,\cite{weinberg}. While we cannot solve this problem at
this stage, the dynamical nature of the DE makes allowance for this
possibility. Furthermore, since there is no obvious gain in the
quintessence idea, we stick to the CC approach, although we extend
it to include the possibility of a dynamical (``running'') $\CC$
term\,\cite{JHEPCC1,SSRev}. The obvious question now is: where this
dynamics could come from?

One possibility is that it could originate from the fundamental
mechanism of inflation\,\cite{inflation}, which presumably took
place in the very early Universe and could have left some loose end
or remnant -- kind of ``fossil'' -- in our late Universe, which we
don't know where to fit in now. However, what mechanism of inflation
could possibly do that? There is in principle a class of distinct
possibilities, in particular see \cite{Sahni98,Mongan01}, but our
very source of inspiration here is the quantum theory of the
conformal factor, which was extensively developed
in\,\cite{Conformal1}. For a recent discussion, see e.g.
\cite{Conformal2,Conformal3} and references therein. More
specifically, we start from the idea of ``tempered anomaly-induced
inflation'', which was first proposed in \cite{shocom,DESY} (see
also \cite{PST}). It leads essentially to a modified form of the
original Starobinsky model\,\cite{Starobinsky}. In the present
paper, we push forward the possibility that the mechanism that
successively caused, stabilized, slowed down (``tempered'') and
extinguished the fast period of inflation in our remote past could
have left an indelible imprint in the current Universe, namely a
very mild (logarithmically) running Newton's coupling $G$. We show
that, if matter is covariantly conserved, this necessarily implies
an effective renormalization group (RG) running of the
``cosmological constant'' energy density, $\rL=\rL(a)$, which takes
the form of a cubic law of $a^{-1}=1+z$ during the matter dominated
epoch ($a$ being the scale factor and $z$ the cosmological reshift).

\section{Anomalous conformal symmetry in cosmology}
\label{sect:conformal}

Following\,\cite{PSW}, we construct a formulation of the Standard
Model (SM) in curved space-time which possesses dilatation
symmetry\,\cite{Coleman}, and extend it to local conformal
invariance in $d=4$\ \cite{shocom}. The action of the theory must
include conformally invariant kinetic terms and interaction terms.
As for scalars $\varphi$ (e.g. Higgs bosons) we take that their
kinetic terms appear in the combination
$(1/2)\,g^{\mu\nu}\partial_\mu\varphi
\partial_\nu\varphi + (1/2)\xi\, R\varphi^2\,,$ which is well-known
to be conformally invariant for $\xi=1/6$ (after using the
non-trivial local conformal transformation law for the scalar of
curvature $R$). The fermion and gauge boson kinetic terms are also
well-known to be conformally invariant. After the standard set of
conformization prescriptions have been applied, the only
non-invariant terms are the massive ones. To fully conformize this
theory at the classical level, we adhere to the procedure of the
Cosmon Model \cite{PSW}, where one replaces these parameters by
functions of some new auxiliary scalar field $\,\chi$. This field is
a background field, and so within the philosophy of QFT in curved
space-time\,\cite{QFTcurved}, it is not submitted (like the metric
itself) to quantization. For instance, for the scalar and fermion
mass terms in the action we replace
\begin{eqnarray}
&&\int d^4x\sqrt{-g}\,\,m_{\varphi}^2\,\varphi^2\ \rightarrow\ \int
d^4x\sqrt{-g}\,\,\frac{m_{\varphi}^2}{\cMd}\,\varphi^2\,\chi^2
\nonumber\\
&&\int d^4x\sqrt{-g}\,\,m\,{\bar\psi}\psi\ \rightarrow \  \int
d^4x\sqrt{-g}\,\,\frac{m}{\cM}\,{\bar\psi}\psi\,\chi\,,
\label{masses}
\end{eqnarray}
where $\cM$ is an auxiliary mass,  e.g. related to a high energy
scale of spontaneous symmetry breaking of dilatation
symmetry\,\cite{PSW}. We expect $\cM$ in the range of the Grand
Unified Theories (GUT's) or higher: $\cM\gtrsim M_X\sim
10^{16}\,GeV$, but certainly below the Planck scale $M_P\simeq
1.22\times 10^{19}\,GeV$. Moreover, there is the Einstein-Hilbert
(EH) action for gravity itself, $S_{EH}$. With the help of the
background field $\chi$, we can conformize it as follows:
\begin{eqnarray}
 S_{EH}\ \ \ \rightarrow\ \ \  S^*_{EH} =
-\frac{M_P^2}{16\pi\,\cMd}\,\int d^4 x\sqrt{-g}\, \left[\,R\chi^2 +
6\,(\partial \chi)^2\,\right]\,. \label{gravity}
\end{eqnarray}
Notice that the setting $\chi=\cM$ on the conformized action
restores the original EH form, as well as all the terms of the
original SM action. This setting (kind of conformal unitary
gauge\,\cite{shocom}) can actually be understood in a more dynamical
sense within the context of non-linearly realized dilatation
symmetry\,\cite{PSW,Coleman}. Namely, by reparameterizing
$\chi={\cM}\,\exp{\left({\Sigma}/\cM\right)}$, the $\Sigma$-field
just shifts under conformal transformations and behaves as the
Goldstone boson (dilaton) of spontaneously broken dilatation
symmetry at the high scale $\cM$. In this context, the setting
$\chi=\cM$ can be thought of as $\chi$ taking a vacuum expectation
value (VEV), with $\Sigma/\cM\ll 1$ because $\Sigma$ performs small
oscillations around it. The full conformized classical action of the
model becomes invariant under the set of simultaneous
transformations
\begin{eqnarray}
(\chi,\varphi) \to (\chi,\varphi)\,e^{-\alpha}, \,g_{\mu\nu}\to
g_{\mu\nu}\,e^{2\alpha}, \,\psi \to \psi\,e^{-3/2\,\alpha},
\label{conformal}
\end{eqnarray}
for any space-time function $\alpha=\alpha(x)$ and for all scalar
and fermion quantum fields $\varphi$ and $\psi$, including the
background metric and scalar field $\chi$.

In this context, the generalized form of the vacuum action in
renormalizable QFT in curved space-time is\,\cite{shocom}:
$S_{vac}\, =\,S_{EH}^{*}+\, S_{HD}$. Here the first term is the
conformal EH term (\ref{gravity}), whereas the second contains
higher derivatives of the metric and can be expressed in the
conformally invariant fashion
\begin{equation}\label{higher}
S_{HD}\, =\, \int d^4x\sqrt{-g}\, \left\{ a_1 C^2 + a_2 E + a_3
{\nabla^2} R \right\}\,,
\end{equation}
where, $a_{1,2,3}$ are some parameters, $C^2$ is the square of the
Weyl tensor and $E$ is the Gauss-Bonet topological invariant in
$d=4$. The total action is
\begin{eqnarray}
S_t =  S_{matter} + S_{vac} + {\bar \Gamma}\,, \label{totality}
\end{eqnarray}
where the part $S_{matter} + S_{vac} $ is classically conformally
invariant.  However, the one-loop part ${\bar \Gamma}$ is not
conformally invariant and constitutes the anomaly-induced
action\,\cite{Conformal1,Conformal2,Conformal3}. To determine it
explicitly, we follow (actually extend) the standard procedure based
on reparameterizing the background fields $(g_{\mu\nu},\chi)$ with
the help of the conformal factor $\sigma$ and a set of (regular)
reference fields $(\bar{g}_{\mu\nu},\bar{\chi})$, as follows:
\begin{equation}\label{reparam}
 \,{g}_{\mu\nu} = e^{2\sigma} {\bar
g}_{\mu\nu}\,, \ \ \ \ \,\chi =e^{-\sigma} {\bar \chi}\,.
\end{equation}
Through these field redefinitions one can solve the functional
differential equation defining the trace anomaly. In the present
case, it has an extra term (the $f$-term):
\begin{eqnarray}\label{Duff}
\lefteqn{ {<T_\mu^\mu> = -\,\frac{2}{\sqrt{-g}} g_{\mu\nu}
\,\frac{\delta \bar{\Gamma}}{\delta g_{\mu\nu}} +
\frac{1}{\sqrt{-g}}\,\chi\, \frac{\delta\bar{\Gamma}}{\delta
\chi}}}\mbox{\hskip 8cm}\\
\lefteqn{ {= -\, \Big\{\,\, wC^2\,+\,bE\,+\,c{\nabla^2}R
\,+\,\frac{f}{\cMd} \,[R\chi^2+6(\partial \chi)^2]
\,\Big\}\,.}}\mbox{\hskip 8cm}\nonumber
\end{eqnarray}
The one-loop values of the $\beta$-functions $\,w,b\,$ and $\,c\,$
are well established since long time ago\,\cite{Desser} and depend
on the matter content of the model. In particular, $c>0$ is required
for stable inflation\,\cite{Starobinsky}. We provide here the
one-loop coefficient associated to the extra term, with the
following result:
\begin{equation}\label{SdW}
f\,=\,\frac{1}{3\,(4\pi)^2}\sum_{F}\,N_F\,\,m_F^2+\frac{1}{2\,(4\pi)^2}\sum_{V}\,N_V\,\,M_V^2\,,
\end{equation}
where $m_F$ and  $M_V$ are the various (Dirac) fermion and vector
boson masses, respectively, and $N_F$ and $N_V$ are their respective
multiplicities (notice that scalars do not contribute in the
conformal case, $\xi=1/6$). We remark that both types of terms in
(\ref{SdW}) are positive definite, hence we infer the important
result that $f>0$ for all possible quantum matter contributions.
Disregarding a conformally invariant term\,\cite{shocom}, one
arrives at the following solution of Eq.\,(\ref{Duff}) for the
anomaly-induced effective action of the combined background fields
$g_{\mu\nu}$ and $\chi$:
\begin{eqnarray}\label{quantum}
\Gamma_{ind}=\int d^4 x\sqrt{-{\bar g}} \,\{w{\bar C}^2\sigma +
b({\bar E} -\frac23 {\bar \nabla}^2 {\bar R})\sigma + 2
b\,\sigma{\bar \Delta}_4\sigma\\ + \frac{f}{\cMd}\,[{\bar R}{\bar
\chi}^2 + 6(\partial {\bar \chi})^2]\sigma\} -\frac{3c+2b}{36}\int
d^4x\sqrt{-{\bar g}}[{\bar R} - 6({\bar \nabla}\sigma)^2 - 6({\bar
{\nabla}}^2 \sigma)]^2\nonumber\,,
\end{eqnarray}
where
\begin{eqnarray}\label{quantum3}
\Delta_4 = \nabla^4 + 2\,R^{\mu\nu}\nabla_\mu\nabla_\nu -
\frac23\,R{\nabla^2} + \frac13\,(\nabla^\mu R)\nabla_\mu
\end{eqnarray}
is the fourth order, self-adjoint, conformal operator acting on
scalars.

In the cosmological context, the conformal factor $\sigma$ is
related to the scale factor through $\sigma=\ln a(\eta)$, where
$\eta=\int dt/a$ is the conformal time. Furthermore, to better
clarify the impact on the EH sector of the total action
(\ref{totality}), let us substitute (\ref{quantum}) in it and
rewrite the final result in the following compact form:
\begin{eqnarray}\label{label}
S_t=&& S_{matter}-\int d^4 x\, \sqrt{-{\bar g}}\
\frac{M_P^2\,(1-\f\ln a)}{16\pi\,\cMd} \,[\,{\bar R}{\bar \chi}^2+
6\,(\partial {\bar
\chi})^2\,]\nonumber\\
&+&\,{\rm higher\ derivative\ terms}\,,
\end{eqnarray}
where we have defined the dimensionless parameter
\begin{equation}\label{ftilde}
\f = \frac{16\pi f}{M_P^2}
=\,\frac{1}{3\pi}\sum_{F}\,\frac{N_F\,\,m_F^2}{M_P^2}+\frac{1}{2\pi}\sum_{V}\,\frac{N_V\,\,M_V^2}{M_P^2}\,.
\end{equation}
In order to project the standard EH frame (in combination with the
higher derivative terms) we set $\chi$ to its VEV, $\cM$, where
conformal symmetry is spontaneously broken; hence, from
(\ref{reparam}), ${\bar \chi}\ =\cM\, e^{\sigma}=\cM\, a$. In
conformal time, the flat FLRW metric is conformally flat, so we have
$\bar{g}_{\mu\nu}=\eta_{\mu\nu}$ and the terms of $S_{HD}$,
Eq.\,(\ref{higher}), trivially decouple from the conformal factor
dynamics. The equation of motion for the scale factor can be
computed from the functional derivative of (\ref{label}), $\delta
S_t/\delta a(\eta)=0$, upon reverting to the cosmic time. The exact
equation for $a=a(t)$ is a rather complicated, non-linear, 4th order
differential equation. A numerical, and also an (approximate)
analytical, solution is given in \cite{shocom,DESY, PST}. The
essential analytic result can be encapsulated in the ``tempered
anomaly-induced solution'', which takes the elegant form
\begin{equation}\label{TAIS}
a(t)=e^{H_P t}\,e^{-\frac14 H_P^2\f t^2}\,,
\end{equation}
in which the parameter $\f$ is seen to play a fundamental role. Here
the scale $H_P$ defines the ``driving force'' for the
anomaly-induced inflation:
\begin{equation}\label{effH}
H_P=\frac{M_P}{\sqrt{-16\pi b}}\,,\ \ \ \ \ b= -\,\frac{N_S +
11N_{F} + 62N_V}{360\cdot (4\pi)^2}\,.
\end{equation}
where $N_S$, $N_F$, $N_V$ are the number of scalars, Dirac fermions
and vector bosons contributing to the one-loop result. We can see
that, in this context, primordial inflation is fundamentally
associated to the Planck scale and also to the existence of the
$b<0$ coefficient, which emerges as a pure quantum matter effect. In
different models of inflation, one finds different energy sources
that trigger the inflationary mechanism\,\cite{inflation}.

Essential in the structure of the solution (\ref{TAIS}) is the fact
that for $\f\neq 0$ the inflationary process is progressively slowed
down (``tempered''\,\cite{shocom}). Thus, one may judiciously
suspect that, starting from general stable inflation conditions
($c>0$), the early Universe should connect gradually with the
Friedmann-Lema\^\i tre-Robertson-Walker (FLRW) phase. Since $H_P\sim
M_P$, we can estimate from (\ref{TAIS}) that this will occur roughly
after $4/\f$ Planck times ($t_P=1/M_P\sim 10^{-43}$ sec.). For a
typical matter content of a GUT (say, for $N_F+N_V\sim 100-1000$)
and $M_X\gtrsim 10^{16}\,GeV$, one can check that $\f$ is in the
range $10^{-5}-10^{-3}$ and, hence, the inflationary period should
typically stop at around a hundred thousand Planck times, at most,
i.e. at $t\sim 10^{-38}$ sec. This dating of the inflationary epoch
lies in the expected range of most inflationary
models\,\cite{inflation}.

Equation (\ref{label}) suggests that the Newton coupling $G\sim
1/M_P^2$ evolves with the scale factor since $M_P^2\rightarrow M_P^2
(1-\f\ln a)$. Defining  $\tau\equiv -\ln a$, the dimensionless
parameter (\ref{ftilde}) can be interpreted as the coefficient of
the $\beta$-function driving the renormalization group equation
(RGE) for the effective (``running'') Newton's coupling:
\begin{equation}\label{dimlessN2}
\frac{\partial}{\partial\tau}\frac{1}{\bar{G}}=\beta_{G^{-1}}(\bar{G})\,,
\ \ \ \ \ \bar{G}(\tau=0)=G_0\,,
\end{equation}
where $\bar{G}=\bar{G}(G,\tau)$ is a function of the renormalized
coupling $G$ at the scale $\tau$, and $G_0\equiv{1}/{M_P^2}$ is the
current value. At the one loop level,
\begin{equation}\label{G1loop}
\beta_{G^{-1}}^{(1)}=\,\frac{\tilde{f}}{G_0}=
\frac{1}{3\pi}\sum_{F}\,{N_F\,\,m_F^2}+\frac{1}{2\pi}\sum_{V}\,{N_V\,\,M_V^2}\,.
\end{equation}
Being $\f>0$, it follows that the parameter $\bar{G}^{-1}$ is
infrared-free and hence the inverse one, $\bar{G}$ (the running
Newton's coupling), is asymptotically free. This can be seen from
the explicit solution of (\ref{dimlessN2}) at one-loop level:
\begin{equation}\label{Ga}
\bar{G}(a)=\frac{G_0}{1-\f\ln a}=\frac{G_0}{1+\f\ln\mu}\,,
\end{equation}
where we observe that $\bar{G}(\mu)\rightarrow 0$ for $\mu\equiv
1/a\rightarrow\infty$.

\section{Running of $G$ and $\CC$ in the present
Universe}

The logarithmic running of the gravitational coupling (\ref{Ga}) is
controlled by the parameter $\f$. Such slow evolution may appear
nowadays as a kind of ``fossil inertia'', reminiscent of the early
inflationary times. Let us note, however, that the potential
infrared effects on the value of $\f$ could not be taken into
account in the above calculation. Therefore, we don't know the
precise prediction for $\f$ at the present time and, in this sense,
it can be treated as a phenomenological parameter. Requiring that it
should not alter significantly the standard picture, we may arguably
suspect that it is still a small number. We will assume that at low
energies (i.e. in the present Universe) it satisfies $0<\f\ll 1$.
Due to the logarithmic character of the law (\ref{Ga}), the running
of the gravitational coupling should be very mild and virtually
undetectable. We remark, however, that even this minute variation
should be understood at a global cosmological level, not as a local
one.

The conformal anomaly, being a short distance effect associated to
inflation in the early Universe, should not distort the formal
structure of Einstein's equations at very large distances. Thus, we
expect essentially the same low energy gravitational theory in the
present Universe. As already mentioned in the previous section, the
full equation of motion is a fourth order, non-linear, differential
equation. When the inflationary phase has stopped, we must recover
the FLRW Universe in the radiation epoch, and therefore the scale
factor grows approximately as $t^{1/2}$. One finds that all higher
order terms in the aforementioned equation of motion decay as
$1/t^4$ whereas the standard ones decay as
$1/t^2$\,\cite{shocom,DESY, PST}. As a result the effect of the
higher order terms in the present Universe is negligible. Moreover,
the terms which are proportional to $\tilde{f}$ can all be absorbed
in $M_P$ according to $M_P^2\rightarrow M_P^2 (1-\f\ln a)$\,, i.e.,
as in (\ref{Ga}). Therefore, only a tiny renormalization of the
parameter $G$ remains in the infrared epoch as a function of the
scale factor. Does this mean that we cannot get any hint of the
primordial dynamics of the early Universe? Not necessarily so. Let
us consider the possible impact on the cosmological term.

From the above considerations, we may assume that at the present
time the gravitational field equations are Einstein's equations with
a non-vanishing $\CC$ term and a slowly running Newton's coupling
$G$. Let us first confirm that the $\CC$ term must indeed be present
in this framework as a consistency requirement. Modeling the
isotropic Universe as a perfect fluid, we have Einstein's equations
in the form
\begin{eqnarray}
R_{\mu\nu }-\frac{1}{2}g_{\mu\nu }R =
8\pi\,\bar{G}\,\tilde{T}_{\mu\nu}\equiv 8\pi
\bar{G}\left[(\rL-\pmr)g_{\mu\nu}+(\rmr+\pmr) U_{\mu}U_{\nu}\right],
\label{EE}
\end{eqnarray}
where $\rmr$ and $\pmr$ are the matter density and pressure.
Consider now the Bianchi identity satisfied by the Einstein's tensor
on the \textit{l.h.s} of Eq,\,(\ref{EE}). It leads to the following
generalized, covariant, local conservation law:
$\bigtriangledown^{\mu}\,\left(\bar{G}\,\tilde{T}_{\mu\nu}\right)=0$,
where we recall that $\bar{G}$ is not constant in this framework. We
can readily evaluate this law in the FLRW metric. If we project the
$\nu=0$ component of it, we find:
\begin{equation}\label{BD}
\frac{d}{dt}\left[\bar{G}\,\left(\rmr+\rL\right)\right]+\bar{G}\,H\,
\alpha_m\,\rmr=0\,,\ \ \ \ \ \  \alpha_m\equiv 3(1+\omega_m)\,,
\end{equation}
where we have introduced the EOS of matter $\pmr=\wm\rmr$, with
$\wm=0,1/3\ (\amr=3,4)$ for cold matter and relativistic matter
(radiation) respectively.   In the following we adhere to the
canonical assumption that matter is conserved, namely
\begin{equation}\label{matterconserv}
\frac{d\rmr}{dt}+\,\alpha_m\,\rho_m\,H=0\,\ \ \rightarrow \ \
\rmr(a)={\rmr^0}{a}^{-\alpha_m}\,,
\end{equation}
where $\rmr^0\equiv\rmr(a=1)$ is the matter density at the present
time. Substituting (\ref{matterconserv}) in the generalized
conservation law (\ref{BD}), we find
\begin{equation}\label{difCCG}
(\rmr+\rL)\,d{\bar{G}}+\bar{G}\,d{\rL}=0\,.
\end{equation}
Admitting that $\bar{G}$ is variable as in ({\ref{Ga}}), this
differential relation implies $\rL\neq 0$. Moreover, it cannot be
satisfied by a strictly constant $\rL$, unless $\rmr(a)=-\rL$ at all
times, which would of course entail a static Universe! Therefore, we
must have $\rL=\rL(a)$ as well!  In other words, the variable
$\bar{G}$ induces a non-vanishing $\rL$ in our Universe, and the
latter must necessarily be dynamical. To determine $\rL(a)$, let us
substitute ({\ref{Ga}}) and (\ref{matterconserv}) in (\ref{difCCG})
and rearrange terms. The final result is
\begin{equation}\label{diffeq} \frac{d\rL}{da}+
P(a)\rL=Q(a)\,,
\end{equation}
where the functions $P$ and $Q$ read
\begin{equation}\label{PQ}
P(a)=\frac{\f}{a\,(1-\f \ln a)}\,,\ Q(a)= -
\frac{\tilde{f}\,\rmr^0}{a^{\amr+1}\,(1-\f \ln a)}.
\end{equation}
The exact solution can be obtained by quadrature as follows:
\begin{eqnarray}\label{rLa}
\rL(a)=(1-\f \ln a)\,\rLo -\f\,\rmr^0\,(1-\f \ln a)\int_{1}^{a}\,
\frac{dx}{x^{\amr+1}(1-\f \ln x)^2}\,,
\end{eqnarray}
where $\rLo\equiv\rL(a=1)$ is the value of the CC density at
present. The last integral cannot be performed in terms of
elementary functions. However, since we expect $\f\ll 1$, we can
just present the result at leading order in $\f$ as follows (if we
also neglect $\f\ln a\ll 1$):
\begin{equation}\label{rLz}
\rL(z)=\,\rLo +\frac{\f\,\rmr^0}{\amr}\left[(1+z)^{\amr}-1\right]\,,
\end{equation}
where for convenience we have recast the result in terms of the
cosmological redshift, $z=\mu-1=(1-a)/a$. From (\ref{rLz}) we see
that, in the matter dominated epoch ($\amr=3$), the cosmological
term evolves as $\rL(z)=A+B (1+z)^3$, i.e., as an approximate
``affine'' ($A\neq 0$) cubic power law of the redshift. This result
is remarkable and encouraging; it tells us that, despite the
extremely slow logarithmic running of the gravitational coupling
with the redshift, the dark energy (in this case, a running
cosmological term) evolves like an (approximate) power law of the
redshift. The cosmological term, therefore, finally reveals as the
truly detectable ``fossil'' (in this case, a ``fossil energy'') that
emerges from this inflationary-inspired scenario. Even if $\f$ is as
small at present as indicated by the high energy computation
($\f\sim 10^{-5}-10^{-3}$), there is a good chance for testing this
model by considering the sensitivity of the cosmological
perturbations to a running $\rL$, e.g. following the approach
of\,\cite{FSS1}. If, however, $\f\sim 10^{-3}-10^{-2}$, the running
of $\rL$  could already be detected from a dedicated EOS analysis of
the DE, see \,\cite{SS12}.

It is useful to write the corresponding generalized Friedmann's
equation (with vanishing spatial curvature) in this model. The
result is the following:
\begin{eqnarray}\label{GFElz}
H^2(a)=\frac{8\pi}{3}\,\frac{G_0}{1-\f \ln
a}\,\left[\frac{\rmr^0}{a^{\amr}}+\rL(a)\right]\,,
\end{eqnarray}
where $\rL(a)$ is given by (\ref{rLa}). To fully analyze the
cosmological consequences of this model, one has to cope with this
complete formula\,\cite{JS2}. However, to order $\f$, and
considering cosmological epochs not very far in the future
(therefore, neglecting again $\f\ln a\ll 1$), it boils down to
\begin{equation}\label{GFEz}
H^2(z)=\frac{8\pi\,G_0}{3}\,\,\left[\rmr^0\,(1+z)^{\amr}+\rL(z)\right]\,,
\end{equation}
with $\rL(z)$ given by (\ref{rLz}). Using  Eq.\,(\ref{GFEz}) and
working within the same approximation, we may rewrite (\ref{rLz}) as
\begin{equation}\label{rLzMDE2}
\rL(z)=\rLo
+\frac{\f}{\amr}\,\frac{3\,M_P^2}{8\pi}\left[H^2(z)-H_0^2\right]\,,
\end{equation}
where $H_0\equiv H(z=0)$. This equation is formally identical to the
one obtained in \cite{RGTypeIa,IRGA03,SSS1} where the running scale
$\mu=H$ was assumed, instead of $\mu=1/a$. The parameter $\nu$
introduced in these references can be identified here with
$\f/\amr$. This correspondence allows us to immediately transfer the
primordial nucleosynthesis bounds on the parameter $\nu$, obtained
in \cite{SSS1} for a $G$-running model similar to the present one,
to the parameter $\f$. This result implies that $\f$ cannot be
larger than $10^{-2}$. In the next section, we further explore the
interesting connections with previous frameworks.

\section{Tracking the running of the cosmological parameters physically.}
\label{sect:runningscales}

Although we have found that the scale factor (or equivalently,
$\mu=1/a$) is the original running scale appearing in our framework,
it is useful to investigate if there are other relevant running
scales, with more physical meaning than $a$, that could be useful to
track the evolution of the cosmological parameters. In the previous
section, we have already mentioned the energy scale defined by the
Hubble function, $\mu=H$. This scale was originally proposed in
\cite{JHEPCC1} and further exploited in \cite{RGTypeIa,IRGA03,SSS1},
see also\,\cite{croat}. Consider now the periods of the cosmic
history more accessible to our observations, i.e. the matter and
radiation dominated epochs. In this case, the approximate formula
(\ref{rLz}) applies to within very good accuracy. Then, from
(\ref{GFEz}), we have
\begin{eqnarray}\label{lnH}
\ln\frac{H^2(z)}{H_0^2}
 \simeq - \amr\,\ln a\,,
\end{eqnarray}
where we assume that $z$ is sufficiently high such that the CC is
subdominant (recall that $\f\ll 1$). The meaning of equation
(\ref{lnH}) is that, in virtually all our observable past, the
running of $G$ in terms of the scale factor can be traced by a
useful physical observable:  the expansion rate $H$. In this way,
the effective coupling $\bar{G}$ in (\ref{Ga}) can approximately be
rewritten as a running function of $H$:
\begin{equation}\label{GalnH}
G(H/H_0)=\frac{G_0}{1+(\f/\amr)\ln (H^2/H_0^2)}\,.
\end{equation}
This result is encouraging because it nicely fits with the previous
result obtained in the alternative framework of \cite{SSS1}, that
is, provided we use (again) the correspondence $\f\leftrightarrow
\amr\nu$ between the basic parameters of the two frameworks. At the
same time, we have the running of the CC term as a function of the
expansion rate:
\begin{eqnarray}\label{rLlnH2}
\rL(H/H_0)=\rLo
+\frac{\f\,\rmr^0}{3}\left(\frac{H^2(z)}{H_0^2}-1\right)\,.
\end{eqnarray}
As it is patent from this equation, the running of the CC term can
be traced by $H$ during the entire matter and radiation dominated
epoch up to the present day. This includes, in particular, the full
range of the supernovae observations.

From the foregoing, we see that $\mu=H$ acts as an alternative
running scale that tracks the evolution of the cosmological
parameters $G$ and $\rL$. Although the primary evolution of these
parameters is formulated in terms of the scale factor, the latter is
not physically measurable. In contrast, the expansion rate $H$ is a
physical observable, which we are measuring nowadays with an
increasing level of precision. In this sense, the evolution of the
cosmological parameters can be better traced through the evolution
of $H$, whenever possible. Furthermore, the use of $H$ as a running
scale allows the present cosmological model to naturally connect
with the previous RG formulations \cite{RGTypeIa,IRGA03,SSS1,SS12}
and, at the same time, to benefit from the various phenomenological
opportunities described there to identify the dark energy as a
dynamical cosmological term. Thus, an advantage of the present
approach is that it preserves essentially all the nice features of
the previous ones while suggesting a potential connection with the
primordial physics of the early Universe. It is in this sense that
the DE that we have detected in our old Universe could be viewed as
a ``fossil'' of the very early times; in fact a ``quantum fossil''
because the non-zero value of the coefficient $\f$ is related to the
quantum effects of matter particles, see Eq.\,(\ref{ftilde}).

Let us note that the trading of $a$ for $H$ ceases to hold when the
Universe becomes highly dominated by the CC since, then, the
Universe is essentially in the de Sitter phase, which means that $H$
becomes constant even though the scale factor starts to grow
exponentially. Clearly, in such circumstances $H$ is not a good
tracer of $a$. Therefore, in the future, when the control of the
evolution is overtaken by an approximately constant cosmological
term $\CCo$, the expansion rate takes the value
$H\simeq\sqrt{\Lambda_0/3}=H_0\,\sqrt{\OLo}\equiv H_{*}$. While this
regime persists, we have $\ln a= H_{*}\,t$ and the original running
law (\ref{Ga}) cannot be mimicked as in (\ref{GalnH}), but as
follows:
\begin{equation}\label{GalnH1}
G(t)\simeq\frac{G_0}{1-\f H_{*}\,t}\,.
\end{equation}
During the quasi-de Sitter regime, the physical scale parameter
tracking the running of $G$ is the cosmic time; equivalently, the
energy scale is $\mu=1/t$.  In this case, $H$ and $\rL$ remain
essentially constant whereas $G$ increases with $t$ as indicated
above. However, this situation will not last forever; one can show
from the full structure of the expansion rate (\ref{GFElz}) -- with
$\rL(a)$ given by (\ref{rLa}) -- and from the equation for
$\ddot{a}$ (the acceleration), that there is a remote future instant
of the cosmic where the Universe will arrive at a turning point in
its evolution. We shall not dwell here on the details of that remote
future epoch, see \cite{JS2}. It suffices to say that the Universe
somehow will recreate in the distant future the tempered inflation
process that it underwent in the past, in the sense that the present
and future state of slow inflation will also cease, roughly after
$1/\f\sim 10^4$ Hubble times, viz. when the turning point will be
approached.

\section{Soft decoupling and running of $\rL$ in the present Universe}

It is important to emphasize that, in this framework, the running of
the cosmological term is actually tied to the running of the
gravitational coupling. This can be better seen if we rewrite the
Bianchi identity (\ref{difCCG}) as follows:
\begin{equation}\label{Bianchi2}
\frac{d\rL}{d\tau}=
G\,(\rmr+\rL)\,\frac{d}{d\tau}\left(\frac{1}{G}\right)
=\frac{3}{8\pi}\,H^2\frac{d}{d\tau}\left(\frac{1}{G}\right)\,,
\end{equation}
where in the second equality we have used Friedmann's equation in
the flat case ($G$ being here, of course, $\bar{G}={\bar{G}}(a)$).
As we have discussed in Section \ref{sect:runningscales}, use of the
expansion rate as the running scale is adequate for most practical
purposes at present and, in addition, it enables us to trace the
running of the parameters in terms of a direct physical observable.
Therefore, let us further transform the RGE (\ref{Bianchi2}) in
terms of the more physical running scale $H$ through the relation
(\ref{lnH}). The latter leads to $d\tau=-d\ln a=(2/\amr)\, d\ln H$.
Using the one-loop result for the RGE of $G^{-1}$,
Eq.\,(\ref{dimlessN2})-(\ref{G1loop}), we may express the desired
differential running law for $\rL$ as follows:
\begin{equation}\label{RGELambdaf}
\frac{d\rL}{d\ln H}=\frac{3\,\nu}{4\pi}\,H^2\,M_P^2\,, \ \ \ \ \ \ \
\ \nu\equiv\frac{\f}{\amr}\,.
\end{equation}
This equation can be thought of as the RGE for the CC density and
has exactly the required form that we suggested on different grounds
in previous approaches to the RG evolution of the cosmological term
(see e.g. \cite{RGTypeIa,IRGA03,SSS1}). As a result, we obtain a
possible unified description of the early and late history of the
Universe in RG terms. Moreover, the physical interpretation of $\nu$
in the present framework is physically the same as in the previous
approach, except that here we have found a possible connection of
this parameter with the mechanism of primeval inflation. Indeed,
with the help of (\ref{ftilde}), we can rewrite $\nu$ in
(\ref{RGELambdaf}) as follows:
\begin{equation}\label{nu}
\nu= \frac{1}{12\pi}\frac{M^2}{M_P^2}\
\end{equation}
with
\begin{equation}\label{M2}
M^2=\frac{4}{\amr}\,\sum_{F}\,{N_F\,\,m_F^2}+\frac{6}{\amr}\sum_{V}\,{N_V\,\,M_V^2}
\equiv\sum_i c_i\,M_i^2\,.
\end{equation}
Here $M_i$ are the masses of all the matter particles contributing
in the loops. Equations (\ref{nu}) and (\ref{M2}) adopt the general
form that we postulated for the RGE of the cosmological term in
\cite{JHEPCC1,RGTypeIa,IRGA03,SSS1}. We see, remarkably enough, that
the heaviest particles provide the leading contribution to the
running of $\rL$. This feature is what we called ``soft-decoupling''
in these references, in the sense that the cosmological term
evolution satisfies, in contrast to the other parameters in QFT, a
renormalization group equation that is driven in part by the
heaviest masses $M_i$ and in part by the physical running scale
$\mu=H$. The running law for the CC density, thus, follows a sort of
generalization of the decoupling theorem\,\cite{AC}. This is related
to the fact that $\rL$ is a dimension-4 parameter. Since there
appears no contribution on the \textit{r.h.s.} of (\ref{RGELambdaf})
that is entirely driven by the masses, otherwise it should be of the
type $\sim M_i^4$ -- and hence disastrous from the phenomenological
point of view\,\cite{JHEPCC1}--, the leading effects are of the
mixed form $\sim H^2\,M_i^2$. For fields whose masses are of order
of the Planck mass ($M_i\lesssim M_P$) the contribution at the
present time is of the order $H_0^2\,M_P^2$. Recalling that $H_0\sim
10^{-42}\,GeV$, we find that the value of $H_0^2\,M_P^2$ falls just
in the ballpark of the current value of the CC density, $\rLo\sim
10^{-47}\,GeV^4$. The running of $\rL$ through (\ref{RGELambdaf})
is, therefore, smooth and of the correct order of magnitude. To be
precise, that RGE tells us that the typical variation of $\rL$ as a
function of $H$ is, at any given time in the cosmic history, of the
order of $\rL$ itself. Let us note that the masses $M_i$ could be
substantially smaller than $M_P\sim 10^{19}\,GeV$ and still get a
sizeable effect in the running of the CC. For example, assume that
there is physics just at a GUT scale $M_X$ a few orders of magnitude
below the Planck scale. This would indeed be the case if we assume
that there is a large multiplicity in the number of particles
involved in that GUT scale. For example, take $\f\sim\nu\sim
10^{-4}$, then from(\ref{nu})-(\ref{M2}) we have
\begin{equation}\label{MX}
\sum_i c_i\,M_i^2=12\,\pi\,M_P^2\,\nu\sim 10^{36}\,GeV^2\,.
\end{equation}
If we assume that the number of heavy degrees of freedom, $M_i\sim
M_X$, is of order of a few hundred, it follows that $M_X\sim
10^{16}\,GeV$. In other words, in this case one could entertain the
possibility that the origin of the RG cosmology could bare some
relation to the physics near the typical SUSY-GUT scale.

To summarize this section, we have found that the soft decoupling
terms $\sim H^2\,M_i^2$ are the leading ones determining the running
of $\rL$. We obtain no $\sim M_i^4$ contributions at all. The
solution of (\ref{RGELambdaf}) that satisfies the boundary condition
$\rL(H=H_0)=\rLo$ is just Eq.\,(\ref{rLzMDE2}), as expected. We
emphasize that the previous interpretation is based on assuming that
the computation of $\f$ in Section \ref{sect:conformal} can be
applied to the present time. As we already warned, there might be
infrared effects that could distort this picture, but we have
assumed that $\f$ will remain small and maybe not essentially
different from what we have found. In particular, within a more
physical RG scheme, and on the grounds of the decoupling
theorem\,\cite{AC}, we expect ordinary decoupling corrections on the
\textit{r.h.s.} of (\ref{dimlessN2}) and (\ref{RGELambdaf}), namely
corrections of the form $(\mu/M_i)^n$ ($n>0$). However, if the scale
$\mu=H$ (defining the typical cosmic energy of the FLRW models) is
used for a physical description of the running of the cosmological
parameters, these corrections should be negligible since $H\ll M_i$
for any known and conceivable particle\,\cite{JS2}.

\section{Conclusions}

In this work we have suggested that the presence of dynamical dark
energy (DE) in the current Universe is actually a consistency demand
of Einstein equations under the two assumptions of: i) matter
conservation, and ii) the existence of a period of primordial
inflation in the early Universe, especially when realized as
``tempered anomaly-induced inflation''. Based essentially on the
previous works\,\cite{shocom,DESY,PST} and on the general setting of
the quantum theory of the conformal
factor\,\cite{Conformal1,Conformal2}, we have found that if the
inflationary mechanism is caused by quantum effects on the effective
action of conformal quantum field theory in curved space-time, then
the gravitational coupling $G$ becomes a running quantity of the
scale factor, $G(a)=G_0/(1-\f\ln a)$, $\f$ being the coefficient of
the $\beta$-function for the conformal Newton's coupling. The effect
of this coupling on the inflationary dynamics is to efficiently
``temper'' the regime of stable inflation presumably into the FLRW
regime. The rigorous high energy calculation of $\f$ in QFT in
curved space-time shows that both fermions and bosons produce
non-negative contributions ($\f\geq 0$). As a consequence, $G$
becomes an asymptotically-free coupling of the scale factor.
Intriguingly enough, we have suggested the possibility that this
running might persist in the present Universe and, if so, it could
provide a \textit{raison d'\^etre} for the existence of the
(dynamical) DE, which would appear in the form of running
cosmological vacuum energy $\rL$. In fact, the logarithmic evolution
of $G$ induces a power-law running of $\rL$, which is essentially
driven by the soft-decoupling terms $\sim H^2\,M_i^2$ (hence by the
heaviest particle masses). The result is a Universe effectively
filled with a mildly-dynamical DE, which can be perfectly consistent
with the present observations.

To summarize, from the point of view of the ``RG-cosmology'' under
consideration, the current Universe appears as FLRW-like while still
carrying some slight imprints of important physical processes that
determined the early stages of the cosmic evolution. Most
conspicuously, the smooth dynamics of $G$ and $\rL$ can be thought
of as ``living fossils'' left out of the quantum field theoretical
mechanism that triggered primordial inflation. Remarkably, this
framework fits with previous attempts to describe the
renormalization group running of the cosmological
term\,\cite{JHEPCC1,RGTypeIa,IRGA03,SSS1,croat,SS12,LXCDM12,Bilic07}
and could provide an attractive link between all stages of the
cosmic evolution.  It is reassuring to find that there is a large
class of RG models behaving effectively the same way. Differences
between them could probably be resolved at the level of finer tests,
such as those based on cosmological perturbations and structure
formation. For example, in references\,\cite{FSS1,GOPS} it is shown
that the study of cosmological perturbations within models of
running cosmological constant puts a limit on the amount of running,
which is more or less stringent depending on the peculiarities of
the model. Similarly, a particular study of perturbations would be
required in the present framework (which includes the variation of
both $\Lambda$ and $G$) to assess the implications on the parameter
$\f$. This study is beyond the scope of the present work.

\vspace{1cm}

\textit{Acknowledgements}. I am very grateful to Ilya Shapiro for
discussions on different aspects of this work and for the fruitful
collaboration maintained on the cosmological constant ptoblem over
the years. I thank also Ana Pelinson for interesting discussions in
the early stages of this work. The author has been supported in part
by MECYT and FEDER under project 2004-04582-C02-01, and also by
DURSI Generalitat de Catalunya under 2005SGR00564 and the Brazilian
agency FAPEMIG. I am thankful for the warm hospitality at the Dept.
of Physics of the Univ. Federal de Juiz de Fora, where part of this
work was carried out.


\end{document}